\documentclass[12pt]{article}
\usepackage{epsfig,amsfonts,amssymb}
\usepackage{cite}
\input epsf.sty
\def\ZZZ{{\hbox{ Z\kern-1.6mm Z}}}
\def\RRR{{\hbox{ R\kern-2.4mm R}}}
\def\CCC{{\hbox{ C\kern-2.0mm C}}}
\def\zzz{{\hbox{z\kern-1mm z}}}

\newcommand{\qeq}{{\hbox{=\kern-2.3mm ? \kern.5mm }}}
\renewcommand{\qeq}{=}

\newcommand{\GG}{{\cal G}}

\newcommand{\HH}{{\cal H}}

\newcommand{\OO}{{\cal O}}

\newcommand{\wt}{\widetilde}
\newcommand{\wh}{\widehat}

\newcommand{\TT}{{\cal T}}

\newcommand{\be}{\begin{equation}}
\newcommand{\ee}{\end{equation}}
\newcommand{\ben}{\begin{eqnarray}\displaystyle}
\newcommand{\een}{\end{eqnarray}}

\newcommand{\bea}[1]{\begin{eqnarray}\label{#1} }
\newcommand{\eea}{\end{eqnarray}}

\newcommand{\refb}[1]{(\ref{#1})}
\newcommand{\p}{\partial}
\newcommand{\sectiono}[1]{\section{#1}\setcounter{equation}{0}}

\def\one{{\hbox{ 1\kern-.8mm l}}}
\def\zero{{\hbox{ 0\kern-1.5mm 0}}}

\begin{document}
{}~
{}~
\hfill\vbox{\hbox{hep-th/0703157}
}\break

\vskip .6cm

{\baselineskip20pt
\begin{center}
{\Large \bf
Geometric Tachyon to Universal Open String Tachyon} 

\end{center} }

\vskip .6cm
\medskip

\vspace*{4.0ex}

\centerline{\large \rm
Ashoke Sen}

\vspace*{4.0ex}

\centerline{\large \it Harish-Chandra Research Institute}

\centerline{\large \it  Chhatnag Road, Jhusi,
Allahabad 211019, INDIA}

\vspace*{1.0ex}

\centerline{\it and}

\vspace*{1.0ex}

\centerline{\large \it Department of Physics, California Institute
of Technology}

\centerline{\large \it Pasadena, CA91125, USA}

\vspace*{1.0ex}

\centerline{E-mail: sen@mri.ernet.in, ashokesen1999@gmail.com}

\vspace*{5.0ex}

\centerline{\bf Abstract} \bigskip

A
system of $k$ Neveu-Schwarz (NS) 
5-branes of type II string theory with one
transverse direction compactified on a circle admits various unstable
D-brane systems, -- some with geometric instability arising out
of being placed at a point of unstable equilibrium in space and some
with the usual open string tachyonic instability but no geometric
instability. 
We discuss the effect of NS 5-branes on
the descent relations among these branes and their physical
interpretation in the T-dual ALF spaces.
We argue that if the tachyon potential controlling these
descent relations obeys certain
conditions, then in certain region in the parameter space labelling
the background
the two types of unstable branes become identical via a
second order phase transition, with the
geometric tachyon in one system getting mapped to the 
usual open string
tachyon of the other system. This would provide
a geometric description of
the tachyonic instability of the usual non-BPS Dp-brane in ten
dimensional flat space-time.

\vfill \eject

\baselineskip=18pt

\tableofcontents

\sectiono{Introduction and Summary} \label{s1}
Type IIA and IIB string theories are known to admit unstable
non-BPS D-branes. These D-branes have tachyonic modes obtained
by quantizing open strings living on these branes. The physics of
the tachyonic mode is by now well 
understood\cite{9805170,9808141,0410103}. 
However there is
no clear geometric interpretation of these modes.

Some time ago Kutasov identified a D-brane system with
a different kind of
instability\cite{0405058,0408073}.  This involves $k$
Neveu-Schwarz (NS) 5-branes with a transverse circle,
and BPS Dp-branes with world-volume parallel to the NS 5-branes,
placed as a point on the transverse circle
diametrically opposite to the NS 5-branes.\footnote{Even though it is
a non-supersymmetric configuration, we shall continue to refer to this
D-brane as BPS D-brane in order to distinguish it from the usual
non-BPS D-branes carrying open string tachyonic modes even in
flat space-time.}
At this point
the potential energy density of the D-brane has a saddle point.
As a result this is a point of unstable equilibrium and 
the  Dp-brane has a tachyonic mode associated with the
displacement of the  brane along the circle.
Although this is a geometric mode, it was found in the analysis of
\cite{0405058,0408073} that the behaviour of this geometric tachyon
is in many ways very similar to the behaviour of the open string
tachyon on a conventional non-BPS D-brane in flat space-time
background. Various other aspects of the dynamics of
this system have been
investigated in 
\cite{0411014,0412022,0501176,0501192,0504226,0510112}.

In this paper we shall introduce several other unstable D-brane systems
in the same background geometry and study and compare 
their properties. 
These additional D-branes are non-BPS D(p+1)-branes extending along
the transverse circle -- either wrapping the circle or ending on the
NS 5-branes -- and other world-volume directions parallel to those
of the Dp-brane described in the last paragraph.\footnote{This
system is different from the one analyzed in 
\cite{0411014} where non-BPS
Dp-branes with world-volume transverse to the circle $S^1$ was
considered.}
These D-branes are unstable due to the usual open string tachyon
living on their world-volume; however they do not have any
additional geometrical instability.

A T-duality transformation on the transverse circle takes the original
closed string field configuration to  
type IIB/IIA
string theory 
on ALF spaces\cite{9511164,9512145}. 
When the $k$ NS 5-branes are all coincident
in the original description, the T-dual geometry involves ALF spaces
with $A_{k-1}$ singularities but when the NS 5-branes are separated
the singularity in the T-dual geometry is resolved by blowing up
the collapsed 2-cycles. The duality map relates the 
geometrically 
unstable Dp-branes in the original background to BPS D(p+1)-branes
wrapped on the equators of these blown up 2-cycles, and the non-BPS
D(p+1)-branes of the original background to
non-BPS Dp-branes and non-BPS D(p+2)-branes wrapped on these
2-cycles. By following the duality map we can derive various properties
of  these branes. For example, one such property
is that the tension of a BPS D(p+1)-brane wrapped on the
equator of a 2-cycle remains finite even when the 2-cycle collapses
to zero size if the D-brane carries a non-zero Wilson line along the
collapsed cycle.

It was already noted in \cite{0405058,0408073} that the geometric
tachyons have many properties in common with conventional open
string tachyons. Since we now
have both types of D-branes in the same
background geometry, we can try to compare their properties in
detail and explore if this analogy can be made into an
equivalence.  Indeed one finds that the two types of D-branes
exihibit very similar behaviour. For example in the NS 5-brane
background the condensation of the
geometric tachyon on the geometrcally unstable BPS Dp-brane and
the condensation of the open string tachyon on the non-BPS
D(p+1)-brane, -- either into the vacuum or into a kink solution 
that depends on any of the $p$ coordinates 
common to both D-branes --
produces identical configurations. Furthermore 
in the absence of the
NS 5-branes the non-BPS D(p+1)-branes and the BPS Dp-branes are
related via usual open string tachyon 
condensation\cite{9812031}.
This continues to hold even in the presence of NS 5-branes, although
we find that the precise form of these relations are
modified. Thus these two types of D-branes may be considered  as
two different classical solutions in the same theory, -- the
world-volume theory of the non-BPS D(p+1)-brane.
This leads to the following question: Can these two solutions merge
as we vary the external parameters {\it e.g.} the radius of the
circle or the number of NS 5-branes?
If so, then at that point the two systems will become identical, with
the geometric tachyon on one getting identified with the
open string tachyon on the other.
Such merger of solutions has indeed been observed in a closely
related system earlier\cite{9906109}.

This analysis however is plagued by the difficulty that the
various properties of the systems with geometric tachyons  like the
tension or the tachyon mass$^2$ were 
calculated using the Dirac-Born-Infeld
(DBI) action and can in principle be affected by
$\alpha'$ corrections.\footnote{We show that the non-BPS
D(p+1)-branes extending along the transverse circle do not suffer
from this problem.} 
When the number of NS 5-branes is large, 
the $\alpha'$ corrections
remain under control even up to the zero radius limit.
In this case we find that
the two solutions do not merge. The BPS Bp-brane in unstable
equilibrium remains lighter than the non-BPS D(p+1)-brane
wrapped on the circle all the way from infinite radius to zero
radius, and  a tachyonic kink configuration on the latter takes us
to the former configuration.
A naive comparison without taking into account $\alpha'$-corrections
shows that in the particular case involving two coincident NS 5-branes
and zero radius of the circle transverse to the NS 5-brane,
the D-brane with geometric instability and the usual non-BPS 
D-brane have identical tension and tachyon mass$^2$, -- a
coincidence already noted in \cite{0408073}. 
In the absence of a non-renormalization
theorem we cannot reach a definite conclusion.\footnote{In a
closely related situation where the NS 
5-brane has non-compact transverse
directions, it is known that the DBI action produces certain results
exactly\cite{0509170}.}
However if it turns out that
there is an underlying non-renormalization
theorem protecting the tension and tachyon mass$^2$ of the system
with geometric tachyon, then it would be a strong indication
that the solutions describing the two types
of
D-branes merge at this point, with 
the geometric
tachyon of one system getting mapped to the usual open string tachyon of
the other system. 

In the T-dual geometry the non-BPS D-brane
under consideration is a non-BPS Dp-brane 
placed at a point
in the ALE space with $A_1$ singularity.
This might lead one to conclude that 
this correspondence, even if true, is not so exciting. 
However we should recall that
the interesting part of the usual open string 
tachyon dynamics is universal and does
not depend on the geometry of the transverse space
in which the D-brane is placed.
In this particular example, the  tachyon vacuum solution,
the kink solutions along any of the $p$ 
directions tangential to the
brane, or the rolling tachyon solution on this Dp-brane
are identical to those on a Dp-brane
in flat space-time. Thus this correspondence, if true, would provide
us with a geometric understanding of the 
tachyon dynamics on a non-BPS Dp-brane
in flat space-time as well.

An interesting case is that of a single
NS 5-brane background with a transverse circle of radius $R$. 
Its T-dual geometry is
Taub-NUT space of size $\wt R=1/R$, which reduces to flat
space-time in the infinite $\wt R$ 
limit. In the original NS 5-brane background we can construct
a geometrically unstable 
D-brane configuration by placing a BPS
Dp-brane at a point
on the transverse circle diametrically
opposite to the  NS 5-brane. 
This configuration should exist for large radius of the transverse
circle, and the
interesting question is: what happens to this brane in the small
radius limit when the dual geometry is flat space-time? 
For multiple NS 5-branes the dual geometry has (collapsed)
2-cycles and the D-brane described above 
has a natural description
as BPS D(p+1)-branes wrapped on the equator of such a 2-cycle. 
However
the geometry dual to a single NS 5-brane
does not possess a 2-cycle, and hence there is no
analogous interpretation for this D-brane.
One can consider several 
possibilities: 1) it could describe a new unstable Dp-brane, 2) it
could disappear from the spectrum by having either zero or
infinite tension, 3) instead of remaining localized,
it could blow up and fill the whole space
in this limit or 4) it could describe the usual unstable Dp-brane.
Of these the fourth
possibility is most exciting, since
it will provide us with a direct
geometric interpretation of the usual open string tachyon on an
unstable Dp-brane in flat space-time as a geometric instability in
the dual description. 

The issue involved is of course the same issue raised
earlier in the more general context of multiple NS 5-brane background,
but it will be useful to describe it again in this special context.
The system of one NS 5-brane with a transverse circle 
contains
a non-BPS D(p+1)-brane wrapped on the transverse circle. 
Its dual description
is the usual non-BPS Dp-brane sitting at the center of the Taub-NUT 
space, -- precisely the configuration with which we would like to
identify the system described in the last paragraph.
Furthermore for large radius of the circle transverse to the five brane,
a kink solution of the open string tachyon on this 
non-BPS D(p+1)-brane can produce
a BPS Dp-brane sitting diametrically across
the NS 5-brane, -- the earlier system. 
Thus two systems can be viewed as different classical
solutions in the same theory, and we would like to ask if the two
solutions can merge at some critical radius
as we reduce the radius of the transverse
circle. If they do then it would mean that below that critical radius the
two solutions become identical. In the dual description it would
imply that the `new' non-BPS Dp-brane constructed via the procedure
described in the previous paragraph 
and the usual non-BPS Dp-brane, sitting at the center
of Taub-NUT, will be identical when the size of the Taub-NUT space
exceeds a critical value. 

We can reformulate this problem in terms of 
an effective potential for the tachyonic mode whose condensation
takes us from the non-BPS D(p+1)-brane to the geometrically unstable
BPS
Dp-brane.\footnote{This tachyonic mode
should be distinguished from the tachyon zero mode 
whose condensation
takes us to the tachyon vacuum.} 
In this description
the non-BPS D(p+1)-brane and the BPS Dp-brane represent 
two different extrema of the tachyon
potential. We show that the quadratic term in the effective potential
changes sign at some critical radius $R_c$ of the circle
transverse to the NS 5-brane. If the coefficient
of the quartic term is positive at this critical radius then
the two extrema merge at this critical radius and remain identical
below this radius.
This represents a second order
phase transition for the BPS Dp-brane
at which the spontaneously broken $(-1)^{F_L}$ symmetry is
restored.
On the other hand if the coefficient of the quartic term in the potential
is negative at the critical radius, then the two extrema do not merge
and the D-branes do not become identical. With our present level of
knowledge we cannot determine what really happens; however we show that
the first possibility is more economical since it does not require the
existence of any other extrema of the potential. 
In contrast if the sign of the
quartic term is negative at the critical radius then new extrema of the
potential appear below the critical radius, signalling new D-brane 
configurations.
If these extrema survive down to zero radius then we have
the problem of explaining what they are in the dual flat space-time
geometry.

This effective potential approach can also be applied for
other values of $k$, but the potential will have different features
for different $k$.
If it turns out that 
there is a non-renormalization
theorem for the tension and the tachyon mass$^2$ of the
geometrically unstable D-branes for $k\ge 2$, then the picture
is somewhat trivial for $k\ge 2$ coincident NS 5-branes. 
For $k=2$ the critical radius where the
two types of D-branes become identical is at $R=0$,
whereas for $k\ge 3$ the two types of D-branes remain distinct
all the way down to $R=0$. 
However the picture becomes much richer once we
consider a more general configuration where the NS 5-branes
are separated from each other.
After all, if we
consider the configuration of $k$ NS 5-branes equally spaced
on a circle of radius $R$, then it is a $k$-fold cover of the
configuration describing a single NS 5-brane with a transverse
circle of radius $R/k$. Thus the merger of the two D-brane
configurations for $R\le R_c$ in the $k=1$ case
will imply that for $k$ equally spaced NS 5-branes,
a BPS Dp-brane sitting midway 
between two neighbouring NS 5-branes and a non-BPS D(p+1)-brane
stretched between the two neighbouring NS 5-branes 
 must become identical below
the critical value $kR_c$ of $R$. In the full moduli space spanned
by $R$ and the separation between the NS 5-branes we would expect a
codimension one critical surface that separates the region in which the
two types of D-branes are distinct from the region in which they are
identical. In the special case of
$k=2$ if we denote by $2c$ the angular separation between the two NS
5-branes on the circle then the critical curve in the $(R,c)$ plane
should pass through the points $(0,0)$ and $(2R_c, \pi/2)$.

If the non-renormalization theorems do not hold then
the detailed picture described above will not be correct, {\it e.g.} for
$k=2$ the critical curve will not pass through the $(0,0)$ point. However
the general picture, \i.e.\ the existence of a critical surface that
separates a region where the two D-branes are identical from the
region where they are not identical, is based on the sign of certain
coefficient in the tachyon potential, 
and will still hold
if this coefficient has the correct sign.

Finally we should remark that even though we have dealt with
unstable systems with tachyons, these may also be useful in getting
stable non-supersymmetric configurations after certain orbifolding
that projects out the tachyon mode. 

The rest of the paper is organized as follows. 
In \S\ref{s2} we describe
various unstable D-brane systems in a 
background of multiple NS 5-branes with a transverse circle and
discuss descent relation between these different D-branes for
large radius of the transverse circle.
In \S\ref{s3} we discuss the description of these unstable D-branes
in the dual ALF geometry.  In \S\ref{s4} we describe the zero
radius limit of the original configuration that converts the dual
ALF
geometry to ALE geometry and study the fate of the descent
relations in this limit. \S\ref{s6} describes comparison
between different D-brane systems and possible identification of
a BPS D-brane with geometric tachyon with a 
non-BPS D-brane with the usual
open string tachyon. In \S\ref{s5} we discuss the case of a
single NS 5-brane with a transverse circle, and determine under
what condition a geometrically unstable D-brane in this 
background  in the
zero radius limit would describe the usual unstable D-brane in a dual
flat space-time geometry.
We conclude in \S\ref{s7} with some comments.

\sectiono {Unstable D-brane Configurations
and Their Descent Relations
in NS 5-brane Background} \label{s2}
We begin by considering a system of $k$ NS 5-branes in type IIA/IIB
string theory stretched along the $(x^0, \ldots x^5)$ plane and placed
at 
$(x^6,\ldots x^9)=(0,\ldots 0)$. Let $x^6$ be a
compact coordinate with period $2\pi R$.
The string metric, the dilaton $\Phi$, and the NS sector 3-form field
strength $\HH$ produced by this background
are given by\cite{9112030,9211056}\footnote{We shall use 
$\alpha'=1$ convention throughout this paper.}
\ben \label{e1}
ds^2 &=& \eta_{\mu\nu}\, dx^\mu dx^\nu
+ H(\vec r, y) (dy^2 + d\vec r^2)\, , \nonumber \\
e^{2\Phi} &=& g^2 \, H (\vec r, y)\, , \nonumber \\
\HH_{mnp} &=& -\epsilon_{mnpq}\p^q \Phi\, , 
\een
where $\mu,\nu$ run from 0 to 5, $m,n,p,q$ run from 6 to 9,
\be \label{e2}
\vec r \equiv (x^7,x^8,x^9), \qquad y \equiv x^6 \, ,
\ee
and
\be \label{e3}
H(\vec r, y) = 1 + {k \over 2 R r} \, {\sinh(r/R)\over \cosh(r/R)
- \cos(y/R)}\, , \qquad r \equiv |\vec r|\, .
\ee
This background is invariant under the
transformation
\be \label{ez2}
\sigma: \quad y\to -y, \qquad \vec r \to -\vec r\, .
\ee
This symmetry will play an important role in our analysis.

If the $k$ 5-branes are not coincident but are placed at
different points $(\vec r_i, y_i)$ ($1\le i\le k$) then the 
solution is still described by \refb{e1}, but with $H$ given by
\be \label{e3xx}
H(\vec r, y) = 1 + \sum_{i=1}^k
{1 \over 2 R |\vec r - \vec r_i|} \, {\sinh(|\vec r - \vec r_i|/R)
\over \cosh(|\vec r - \vec r_i|/R)
- \cos\left((y-y_i)/R\right)}\, .
\ee

We shall consider various types of non-supersymmetric D-brane
configurations in the background geometry described in
eqs.\refb{e1}-\refb{e3}. The first type of
such configurations, which we shall call G-type D-branes because they
will turn out to have {\bf g}eometric instability, is obtained by
placing a BPS Dp-brane ($p\le 5$) along 
$(x^0,\ldots x^p)$ at
$\vec r=\vec 0$, $y=\pi R$ and arbitrary values
of $x^{p+1},\ldots x^5$\cite{0408073}.  We shall summarize
the main results of \cite{0408073}.
The DBI action on the Dp-brane in this
background will be given by
\be \label{e4}
-g^{-1}\, 
\TT_p\, \int d^{p+1} \xi \, 
\left(H(\vec Z, Y)\right)^{-1/2}\, \sqrt{-\det G} \, 
\ee
where $\{\xi^\alpha\}$ ($0\le\alpha\le p$) are the
Dp-brane world-volume coordinates -- taken to coincide with
$(x^0,\ldots x^p)$, $\vec Z$ and $Y$ denote 
respectively the $\vec r$ and $y$ coordinates of the D-brane 
world-volume,
$g^{-1}\,
\TT_p$ denotes the tension of a BPS Dp-brane at $\infty$,
and 
\be \label{e5}
G_{\alpha\beta} = \eta_{\alpha\beta} + H(\vec Z, Y)
(\p_\alpha \vec Z \cdot \p_\beta\vec Z + \p_\alpha Y \p_\beta Y)\, ,
\qquad 0\le\alpha,\beta\le p\, ,
\ee
is the induced metric on the Dp-brane world-volume.
Note that we have ignored the motion of the brane along
the plane of the 5-brane as well as the dynamics of gauge
fields on the brane; these will not play any role in our analysis.

The overall multiplicative factor of $(H(\vec Z, Y))^{-1/2}$ provides
a potential for the motion of the brane.  For $H$ given in
\refb{e3} this potential has an absolute minimum
at $Y=0$, $\vec Z =0$ where it vanishes, and a saddle point at
$Y=\pi R$, $\vec Z =0$ where it has a minimum as a function of
$\vec Z$ but a maximum as a function of $Y$. Thus it represents
a point of unstable equilibrium\cite{0408073} and $Y$
becomes a tachyonic field on the Dp-brane world-volume. 
One can easily calculate the tension
$\tau_p$ of the Dp-brane situated at $(\vec Z=\vec
0, Y=\pi R)$ and the
mass squared $m_T^2$ of the tachyon on this D-brane coming from
the unstable geometric mode. The answers are
\ben \label{e6}
\tau_p &=& g^{-1} \TT_p H(\vec 0,\pi R)^{-1/2}
\qeq g^{-1}\, \TT_p \, \left( 1 + {k\over 4 R^2}\right)^{-1/2}\, ,
\nonumber \\
m_T^2 &=& \left[ H^{-1/2} {\p^2 H^{-1/2}
\over \p y^2}\right]_{\vec r=0,
y=\pi R} \qeq -{k\over (k + 4R^2)^2}\, .
\een
We should keep in mind however that 
the DBI action \refb{e4} 
receives higher derivative corrections. Thus
the results \refb{e6} can also 
receive higher derivative corrections. In particular for low values
of $k$ and $R\sim 1$,
the spatial curvature as well as the derivatives of
the dilaton are of order one
near $(\vec r=0, y=\pi R)$ and hence the corrections can be of order
unity. We shall return to a discussion about these corrections
later.

It was noted in \cite{0408073} that the tachyonic mode described
by $Y$ has many properties in common with the usual
open string tachyon on an
unstable D-brane.
Clearly the minimum of the tachyon potential is at $Y=0$ where
the tension of the Dp-brane, being proportional to
$H(\vec 0,0)^{-1/2}=0$,  vanishes. 
Furthermore one can consider tachyonic
kink configurations in this theory located at $x^p=0$, described by
the solution $Y=0$ for $x^p<0$ and $Y=2\pi R$
for $x^p>0$. This describes a BPS Dp-brane located at $x^p=0$
and stretched along $x^0,\ldots x^{p-1}$, $y$\cite{0408073}.
Neither of these general properties is expected to be modified by
$\alpha'$ corrections.

For future reference it will be useful to consider the situation
where the $k$ NS 5-branes are displaced away from the $(\vec r, y)
=(\vec 0, 0)$ point symmetrically to
$\{ (\vec r_i, y_i), \, 1\le i\le k\}$ 
so that $(\vec r, y)=(\vec 0, \pi R)$
is still a point of unstable equilibrium for the Dp-brane. In this case
\refb{e3xx} shows that
the tension of the Dp-brane situated at $(\vec 0, \pi R)$ is given
by
\be \label{etenag}
\tau_p = g^{-1} \, \TT_p \, H(\vec 0,\pi R)^{-1/2}
\qeq g^{-1}\, \TT_p \, \left(
1 + \sum_{i=1}^k
{1 \over 2 R |\vec r_i|} \, {\sinh(|\vec r_i|/R)
\over \cosh(|\vec r_i|/R)
+ \cos\left(y_i/R\right)}
\right)^{-1/2}\, .
\ee

A second type of non-supersymmetric D-brane configuration in the
same background is obtained by placing
a non-BPS D(p+1) brane\cite{9805170,9808141,0410103,9806155}
spanning the coordinates $x^0,\ldots x^p$ and $y$.
We shall call these U-type D-branes to indicate that they carry the
{\bf u}sual open string tachyon.
The world-volume
action describing the dynamics of the massless modes on
this brane is given by
\be \label{en1}
-{1\over \sqrt 2 \pi}\, g^{-1} \, \TT_p\, 
\int d^{p+1}\xi \, dy\, H(\vec Z, y)^{-1/2}\, \sqrt{-\det \GG}\, ,
\ee 
where $(\{\xi^\alpha\}, y)$ denote the world-volume coordinates and
\be \label{en2}
\GG_{\alpha\beta} = \eta_{\alpha\beta} + H(\vec Z, y)
\p_\alpha \vec Z \cdot \p_\beta\vec Z, \quad \GG_{yy} = 
H(\vec Z, y) \left(1 + \p_y \vec Z \cdot \p_y\vec Z\right)
\, , \quad
\GG_{\alpha y}=\GG_{y\alpha} = H(\vec Z, y) \p_\alpha\vec Z
\cdot \p_y \vec Z\, .
\ee
As before we have ignored the dynamics of the gauge fields on the
brane and the motion of the brane along the NS 5-brane world-volume.
Another field that has not been included in the action \refb{en1} but
will play an important role in our analysis is the open string tachyon
field.

This theory has a classical solution corresponding to
\be \label{en3}
\vec Z = \vec c\, ,
\ee
for any constant vector $\vec c$, 
describing the non-BPS D(p+1)-brane situated as $\vec r=\vec c$ and
spanning the $x^0,\ldots x^p,y$ directions. 
{}From the
point of view of an asymptotic observer this looks like a p-brane
since one of its world-volume directions is compact.
The tension, defined as the mass per unit $p$-volume,
and the mass$^2$ of the open string tachyon living
on this D-brane is
given by
\be \label{en5}
\tau_p' = {1\over \sqrt 2 \pi}\, g^{-1} \, \TT_p\, 
\int dy  \, H(\vec Z, y)^{-1/2}\, 
\sqrt{-\det \GG}\bigg|_{\vec Z=\vec c}
= \sqrt 2\, R\, g^{-1} \, \TT_p\,
\qquad m_T^{\prime 2} = -{1\over 2}\, .
\ee
This D-brane
can decay into the closed string vacuum of zero energy density
via tachyon 
condensation\cite{0001084,0002211,0003220,0004015}.
Furthermore an $x^p$ dependent 
tachyonic kink solution on this D-brane localized
along the $x^p=0$ surface produces a BPS Dp-brane stretched along
the $x^0,\ldots x^{p-1}$, $y$ direction\cite{9812031}. 
Due to the universality of the open string tachyon 
dynamics\cite{9911116,0005036} neither of these properties
is modified by $\alpha'$ corrections.
Thus the condensation of the open string tachyon living on
this D-brane, either into a kink or into its vacuum,
produces identical results as the condensation of the geometric
tachyon living on the BPS Dp-brane placed at $y=\pi R$.
This provides us with the first hint that there may be some
deeper relation between these branes.

We shall now argue that unlike the formul\ae\ 
\refb{e6}, eqs.\refb{en5} are not modified by $\alpha'$
correction. We begin with the tension. 
For this let us switch theories and consider type IIB/IIA theory
with the same NS5 background and a BPS D(p+1) brane along
$x^0,\ldots x^p,y$. 
This is a supersymmetric system with no force between
the D-brane and the NS 5-brane and hence the tension of the
D(p+1)-brane (defined as the mass per unit $p$-volume after
integration over the $y$ coordinate)
is given by the BPS formula which is
independent of $\vec c$ and is not modified by $\alpha'$
corrections. The same argument holds for a BPS
$\bar {\rm D}$(p+1)-brane in the same position. {}From this we can
conclude that at string tree level when the interaction between
different D-branes can be ignored, the tension of a coincident
D(p+1)-brane-$\bar{\rm D}$(p+1)-brane at an arbitrary position
$\vec c$ will be equal to 
twice that of a single D(p+1)-brane. We can now take an orbifold
of this system by $(-1)^{F_L}$ to construct a non-BPS D(p+1)-brane
of IIA/IIB placed at $\vec c$\cite{9812031}, 
with its tension given by $\sqrt 2$ times that of the
BPS D(p+1)-brane in IIB/IIA theory. This is precisely the tension
given in \refb{en5}. Thus we conclude that there is
no $\alpha'$ correction to the expression of the tension given in
\refb{en5} for any $\vec c$.

The argument regarding tachyon mass$^2$ is even simpler. Since the
tachyon vertex operator on the non-BPS D-brane is proportional to
the identity operator in the matter sector, and since the identity
operator has dimension zero in all conformal field theories, the
mass$^2$ of the tachyon on the non-BPS D(p+1)-brane must be given by
\refb{en5}. 

A major part of our analysis will focus on studying the relationship
between the two types of D-branes introduced above. We shall begin by
comparing the action of the transformation $\sigma$
given in
\refb{ez2} on the tachyon on both types of branes.
Since $\sigma$ changes $y\to -y$, the geometric tachyon $Y$ on the
G-type D-brane, being the 
$y$ coordinate of the D-brane world-volume, clearly changes sign
under 
$\sigma$. On the other hand it is known from the analysis of
\cite{9812031} that the open string tachyon field $T$ on the
unstable D(p+1) brane wrapped along $y$ also
changes sign under $\sigma$. 
Thus if we restrict ourselves to $\sigma$ invariant
field configurations, we project out the geometric tachyon on the
G-type D-brane as well as the zero mode of the open string
tachyon on the U-type D-brane. However, for sufficiently large
radius of the transverse circle the lowest lying $\sigma$ invariant
mode of the open string
tachyon on the U-type D-brane, satisfying $T(y)=-T(-y)$, 
is also tachyonic, and its condensation
produces a Dp-brane at $y=\pi R$ and a ${\bar {\rm D}}$p-brane at
$y=0$\cite{9812031}. 
Since the ${\bar {\rm D}}$p-brane at
$y=0$, being
on top of the NS 5-branes, has no tension
and charge, it is indistinguishible
from the tachyon vacuum. Thus
the resulting configuration is just  a Dp-brane at $y=\pi R$, \i.e.\
the G-type Dp-brane.  This shows that both G and U-type D-branes
can be regarded as different classical solutions in a single theory
-- the world-volume theory of the U-type brane.
If we reduce  the radius, then in the absence of 5-branes the situation
is reversed at a critical radius, and the U-type D-brane is
obtained as a result of winding tachyon condensation on the
G-type brane\cite{9812031}. 
We shall see in \S\ref{s4}-\S\ref{s5} that the situation changes when
NS 5-branes are present.

There is a third type of non-supersymmetric brane configurations
with properties very similar to those of the U-type branes
described above. 
Here we
consider again a non-BPS D(p+1) brane
spanning the coordinates $x^0,\ldots x^p$ and $y$, but this time
instead of wrapping the $y$ circle 
it begins at one of the $k$ NS5-branes,
goes around the circle, and ends on another NS5-brane. 
We shall call these S-type D-branes to indicate that
they are non-BPS D-branes {\bf s}tretched between the NS 5-branes.
When all the
5-branes are coincident this configuration has the same tension and
the tachyon mass$^2$ as the original configuration; however this
is no longer the case if we separate the five-branes. For example if we
take a pair of 5-branes at $y=y_1$ and $y=2\pi R- y_2$
respectively, then a non-BPS D(p+1)-brane stretching between these
five branes will have tension and tachyon mass$^2$ given by
\be \label{eten1}
\tau_p''=
{1\over \sqrt 2 \pi}\, g^{-1} \, \TT_p \, (2\pi R - y_1 - y_2),
\qquad m_T^{\prime\prime 2} = -{1\over 2}\, .
\ee
The tension coincides with $\tau_p'$ given in 
\refb{en5} for $y_1=y_2=0$
but not otherwise. In particular if we take $y_1=y_2=\pi R$
then the D-brane will have vanishing tension.
Another difference between this D-brane
and the U-type D-brane considered earlier is that the latter
can be moved away from $\vec r=0$ at no cost in
energy, but that is not the case for the new system since the 5-branes
are located at $\vec r=0$. An argument similar to that for the
U-type D-branes shows us that the tension and the tachyon mass
formul\ae\ for the S-type D-brane also 
do not receive any
$\alpha'$ corrections. An $x^p$-dependent 
tachyonic kink on this D-brane localized
at $x^p=0$ will produce
a BPS Dp-brane along $x^0,\ldots x^{p-1},y$, ending on the two
NS 5-branes at the two ends. On the other hand a 
$y$-dependent tachyonic kink
localized at $y=\pi R$ will produce a BPS Dp-brane at $y=\pi R$
lying along $x^0,\ldots x^p$, \i.e.\ a G-type unstable
Dp-brane configuration.

\sectiono {Dual Description in ALF Spaces} \label{s3}
We now consider a different description of the system related to
the one given above by a T-duality transformation
along the circle along $y$.
This maps the closed string backgound involving the NS 5-branes to
a configuration in type IIB/IIA theory
of $k$ coincident Kaluza-Klein monopoles, or
equivalently ALF space\cite{gibhaw} with 
$A_{k-1}$ singularity at the origin\cite{9511164,9512145}.
The dilaton and the metric associated with this background is
given by
\ben\label{tnutgeom}
ds^2 &=& \left(1+\frac{k\wt R}{2r }\right)
\left( dr ^2 +  r ^2 ( d\theta^2 + \sin^2 \theta 
d\phi^2) \right)
+ \wt R^2\left( 1 + \frac{k\wt R}{2r }
\right)^{-1} \left( 
\, d\psi + {k\over 2}\cos\theta d\phi\right)^2 \nonumber \\
e^{2\Phi} &=& \wt g^2\, , 
\een
with the identifications:
\be\label{eident}
(\theta,\phi,\psi) \equiv (2\pi -\theta,\phi+\pi, \psi+{k\pi\over 2})
\equiv (\theta+2\pi, \phi+2\pi,\psi)
\equiv (\theta,\phi+2\pi,\psi+k\pi)\equiv (\theta,\phi,\psi+2\pi)\, .
\ee
Here
\be\label{eg1}
\wt R = {1/  R}, \qquad \wt g^2 = g^2  / R^2\, .
\ee
{}From the point of view of an asymptotic observer $(r,\theta,\phi)$
are the polar coordinates with origin at the location of the monopoles
and $\psi$ is the coordinate along the compact direction.
For $k=1$ the geometry is smooth but for $k>1$ there is an $A_{k-1}$
singularity at the origin $r =0$.
This is best seen by introducing the cartesian coordinate system for
the metric near the origin:
\bea{ecart}
&& w^1 = 2\sqrt r \cos{\theta\over 2}\cos\left(
{\psi\over k}+{\phi\over 2} \right), \qquad w^2 =
2\sqrt r \cos{\theta\over 2}\sin\left(
{\psi\over k}+ {\phi\over 2} \right), \nonumber \\ &&
 w^3 =  2\sqrt r \sin{\theta\over 2}\cos\left({\psi\over k} -
{\phi\over 2}\right),
\qquad w^4 =  2\sqrt r \sin{\theta\over 2}\sin\left({\psi\over k} -
{\phi\over 2}\right)\, .
\eea
For $k=1$ the coordinates $\{w^a\}$ are invariant under the
identification
\refb{eident}, but for $k>1$ there is an identification
under a $\ZZZ_k$ rotation in the $w^1$-$w^2$ and 
$w^3$-$w^4$ plane.

It will be useful to examine the action of the transformation
\refb{ez2} on the dual geometry. This can be figured out by examining
its action at large $|\vec r|$ where the $(\vec r, y)$ space looks
like $\RRR^3\times S^1$. Under a T-duality that takes the circle
labelled by $y$ to its dual circle labelled by $\psi$,
the image of \refb{ez2} is known to be given by
$(-1)^{F_L}$ accompanied by $\vec r\to -\vec r$, $\psi\to -\psi$
transformation. In terms of $(r,\theta,\phi)$ coordinates 
this translates to
\be \label{ez2new}
(-1)^{F_L} \times \left\{
(r,\theta,\phi,\psi)\to (r,\pi-\theta,\phi+\pi,-\psi)\right\}\, .
\ee

The solution described in \refb{tnutgeom} 
is actually a special limit of
a general class of non-singular solutions given by
\ben\label{tnutgen}
ds^2 &=& U(\vec r) 
\left( dr ^2 +  r ^2 ( d\theta^2 + \sin^2 \theta 
d\phi^2) \right)
+ \wt R^2 U(\vec r)^{-1} \left( 
\, d\psi + {1\over \wt R} \vec\omega.d\vec r\right)^2 \nonumber \\
e^{2\Phi} &=& \wt g^2\, , 
\een
where
\be \label{eudef}
U(\vec r) = 1 + {\wt R\over 2}\, 
\sum_{i=1}^k {1\over |\vec r - \vec r_i|},
\qquad \vec\nabla\times \vec\omega = \vec\nabla U\, .
\ee
This space is completely non-singular; the apparent singulaities 
at $\vec r=\vec r_i$ being coordinate singularities.
There are various non-contractible two cycles described
by taking the direct product of the circle labelled
by $\psi$ and the straight line in $\vec r$ space joining $\vec r_i$
and $\vec r_j$. Away from $\vec r_i$ and $\vec r_j$ the resulting two
dimensional surface looks like a cylinder but the circle labelled
by $\psi$ collapses at the end-points $\vec r_i$ and $\vec r_j$ giving
this surface the topology of a sphere. 
In the limit $\vec r_i\to \vec r_j$ the 2-cycle passing through
$\vec r_i$ and $\vec r_j$ collapses and the space becomes singular.
Explicit discussion on D-branes wrapped on various 2-cycles
of this geometry
can be found in \cite{9707123}.
 
This geometry described in \refb{tnutgen}
is dual to the configuration of NS5-branes described
in eq.\refb{e3xx} with all the $y_i$'s set to zero. 
Non-vanishing $y_i$'s correspond to switching on flux of
NS sector 2-form
field B through various 2-cycles of this 
geometry\cite{9511164,9512145}.

We shall now describe the various unstable D-brane configurations
introduced in the previous section in this T-dual background. We
begin with the G-type unstable D-branes obtained in the
original description by placing
a BPS Dp-brane
along $x^0,\ldots x^p$
at $(\vec r=0, y=0)$ or $(\vec r =0, y=\pi R)$.
Since T-duality acting on a D-brane localized at a point on a circle
maps it to a D-brane wrapped along the dual circle, we expect that
the 
T-dual description of the G-type brane is 
a BPS D(p+1) brane along
$x^0,\ldots x^p$ and $\psi$, placed at 
$\vec r = 0$.\footnote{We must caution the reader that this
heuristic picture should be used with caution; $\alpha'$ corrections
necessarily spread out the D-brane wave-function and hence the
D-brane boundary states are not strictly localized either in the
original description or in the new description. However when all
the $|\vec r_i|$'s are large and hence space-time near $\vec r=0$ is
nearly flat, this picture becomes accurate.}
The coordinate $y$ of the original
Dp-brane corresponds to the Wilson line along $\psi$ on the
dual D(p+1)-brane. 
In order to test this we begin with a
configuration where in the original description we move the NS 5-branes
far away from the origin in a symmetric fashion so that $(\vec 0, 0)$
and $(\vec 0, \pi R)$  
 remain extrema of the potential. In this case from 
\refb{etenag} we see that the tension of the Dp-brane is given by
\be \label{etrep}
g^{-1}\, \TT_p \, \left(
1 + \sum_{i=1}^k
{1 \over 2 R |\vec r_i|}\right)^{-1/2} + \OO\left( e^{-|\vec r_i|/R}
\right)\, .
\ee
On the other hand in the dual
description, the mass per unit $p$-volume
of a D(p+1)-brane wrapped along the $\psi$ circle at $\vec r=0$
is obtained by integrating the tension of the D(p+1)-brane along
the $\psi$ circle:
\be \label{etencomp}
\wt g ^{-1} { \TT_p \over 2\pi}\,\int d\psi\, {\wt R} \,
U(\vec r)^{-1/2} = g^{-1}\, \TT_p \,  \left(
1 + \sum_{i=1}^k
{1 \over 2 R |\vec r_i|}\right)^{-1/2}\,.
\ee
This agrees with \refb{etrep} up to exponentially suppressed terms.
At the level of the supergravity approximation
we do not see a potential for the Wilson line or the exponentially
suppressed terms in \refb{etrep}; however these are expected
to be induced by the world-sheet instanton 
corrections\cite{0204186,0507204,0508097}.
Physically the $\psi$-circle at $\vec r=\vec 0$ represents the equator
of a blown up 2-cycle. Thus we see that the T-dual description of the
G-type Dp-brane is a BPS D(p+1)-brane wrapped along the
equator of a 2-cycle.

Once we have made the identification for large $|\vec r_i|$ we can now
take the $\vec r_i\to 0$ limit on both sides. 
Let us focus on the $y=\pi R$ case.
In the original
description this gives a BPS Dp-brane of IIA/IIB
placed diametrically opposite to
a set of $k$ coincident NS 5-branes on a transverse circle.
In the T-dual description the circle at $\vec r=0$
labelled by $\psi$
collapses to a point in this limit and 
we get a BPS D(p+1)-brane of IIB/IIA
wrapped along this collapsed circle, but carrying half a unit of
Wilson line along this circle. According to \refb{etencomp}
the tension of this brane vanishes in this limit. This result however is
likely to be $\alpha'$ corrected for all $R$
since the curvature at $\vec r=0$ is
strong. For small $\wt R$ or
equivalently 
large $R$ we can trust the computation of the brane tension
in the original geometry, and
by exploiting the duality invariance we come to the conclusion that
the tension of this brane remains finite and is given by
\refb{e6} in this limit if we switch on half a unit of Wilson line
on the brane along the $\psi$ direction. On the other hand if the
Wilson line is zero then it corresponds to placing the Dp-brane at
$(\vec r, y)=(\vec 0, 0)$, \i.e.\ at the location of the NS 5-branes,
in the original description. Thus the tension of the brane vanishes
in this case.

The $k=1$ case deserves special attention. In this case there is
a single NS 5-brane in the original description, and we cannot
move the NS brane away from $(\vec 0, 0)$ keeping
$(\vec 0, \pi R)$ a point of unstable equilibrium. Nevertheless by
using the symmetry of the problem we could interpret the
dual of the BPS Dp-brane placed at $(\vec r, y)=
(\vec 0, \pi R)$ as
a BPS D(p+1)-brane wrapped along the $\psi$ circle at $\vec r=0$
with Wilson line along the $\psi$ direction. Again for sufficiently
small $\wt R$, \i.e.\ 
large $R$, this D-brane has finite tension given by \refb{e6}
with $k=1$. This is somewhat surprising considering that the
$\psi$ circle collapses at $\vec r=0$ and
Taub-NUT
space has
no singularity at $\vec r=0$. We shall return to this case in \S\ref{s5}.

The non-BPS D(p+1)-brane of the original theory
along $x^0,\ldots x^p$, $y$ placed at
$\vec r$ are easy to describe in the dual theory. This goes over to
a non-BPS Dp-brane along $x^0,\ldots x^p$ in the dual system placed at
fixed values of $\vec r$ and $\psi$ in the ALF space, with the
location along $\psi$ determined by the Wilson line along $y$ of the
original system. In particular the original D(p+1) brane placed
at $\vec r=0$ corresponds to a non-BPS Dp-brane placed at 
$\vec r=0$ in the
dual system.

Finally we turn to 
the S-type D-brane, extending from the $i$th to the
$j$th NS 5-brane in the original description. Again the simplest way
to find their dual description is to first move the NS 5-branes away from
$(\vec r,y)=(\vec 0,0)$ to positions $(\vec r_i, 0)$. In this process the 
non-BPS D(p+1)-brane extending from the $i$th to the $j$th NS 5-brane
gets stretched between the points $(\vec r_i, 0)$ and $(\vec r_j, 2\pi R)$.
Had the $y$ coordinate been zero at both the end points the D(p+1)-brane
would have been fully localized in the $y$-direction and a T-duality
transformation would have delocalized it along the dual $\psi$
direction, producing a non-BPS D(p+2)-brane wrapped around the
2-cycle passing through $\vec r_i$ and $\vec r_j$. By following the 
standard rules of T-duality transformation one can show that the
effect of one unit of 
winding of the original D(p+1)-brane along the $y$
direction is to produce one unit of gauge field strength flux on the
dual D(p+2)-brane through this 2-cycle. Thus the S-type brane
corresponds in this dual description
to non-BPS D(p+2)-brane wrapped on the two cycle passing through
$\vec r_i$ and $\vec r_j$,
with one unit of gauge field strength flux turned on through this
2-cycle. When
the 5-branes in the original description are coincident, the 2-cycle
collapses to zero size with
vanishing flux of the 
$B$ field through them. In this case the
D(p+2)-brane would have vanishing tension if  it did not have a gauge
field flux turned on. The gauge field flux however makes this into a
non-BPS Dp-brane. Separating the NS 5-branes in the original description
along the $y$ direction
corresponds to switching on $B$-flux
through appropriate 2-cycles in the dual description. 
If we consider the special
situation where the $i$th and the $j$th
NS 5-branes are brought
at $y=\pi R$, it corresponds in the dual
description to  one unit of $B$-flux through the 
corresponding 2-cycle.
In this bulk this is equivalent to having no flux, and hence 
vanishing cycles, but its effect on the D(p+2)-brane is to switch off
the flux of the gauge field strength.
As a result the D-brane would now really have vanishing tension,
in agreement with what happens in the original description in
terms of NS 5-branes.

\sectiono{Small Radius Limit and the Fate of the Descent
Relations} \label{s4}
Let us now take the limit\cite{0408073}
\be \label{elimit}
R\to 0, \qquad g\to 0, \quad \wt g\equiv {g\over R} \quad \hbox{fixed}\,,
\ee
and define new coordinate
\be \label{es1}
\wt y = {y\over R}, \qquad \vec {\wt r} = {\vec r\over R}\, ,
\ee
in the original theory.
In this limit 
eqs.\refb{e1}, \refb{e3} take the form
\be \label{ex1}
ds^2 = \eta_{\mu\nu}\, dx^\mu dx^\nu
+ h(\vec {\wt r}, \wt y) (d\wt y^2 + d\vec {\wt r}^2)\, , 
\qquad
e^{2\Phi} = \wt g^2 \,  h(\vec {\wt r}, \wt y) \, ,
\ee
\be \label{ex3}
h(\vec {\wt r}, \wt y) =  {k \over 2  \wt 
r} \, {\sinh\wt r\over \cosh\wt r
- \cos\wt y}\, , \qquad \wt r \equiv |\vec {\wt r}|\, .
\ee
In this coordinate system the BPS Dp-brane in unstable equilibrium
is situated at $\vec{\wt r}=0$, $\wt y =\pi$.
The formul\ae\ \refb{e6} for the tension and the tachyon
mass$^2$ of this G-type brane take the form\cite{0408073}:
\be \label{e6scale}
\tau_p \qeq {2\over \sqrt k}\, \wt g^{-1}\, \TT_p \, ,
\qquad m_T^2 \qeq -{1\over k}\, .
\ee
On the other hand in this limit eq.\refb{en5}, describing the tension
and tachyon mass$^2$ of the U-type brane, --
a D(p+1)-brane wrapped along
$\wt y$, -- reduces to
\be \label{en5scale}
\tau_p' =  \sqrt 2\, \wt g^{-1} \, \TT_p\, ,
\qquad m_T^{\prime 2} = -{1\over 2}\, .
\ee
Finally the dual ALF space itself reduces to $\CCC^2/\ZZZ_k$ 
in this limit
since the size $\wt R=1/R$ of the space goes to $\infty$.
The string coupling in this dual theory is given by $g/R=\wt g$.

Again \refb{e6scale} can receive $\alpha'$ corrections. For finite
$k$ these corrections can be of order unity. However for
large $k$ eq.\refb{e6scale} still remains a valid approximation. To
see this let us define new coordinates
\be \label{enewc1}
\wh y = \sqrt k\, \wt y, \qquad \vec {\wh r} = \sqrt k \,
\vec{\wt r}\, ,
\ee 
and express \refb{ex1}, \refb{ex3} as
\be \label{ex1new}
ds^2 = \eta_{\mu\nu}\, dx^\mu dx^\nu
+ \wh h(\vec {\wh r}, \wh y) 
(d\wh y^2 + d\vec {\wh r}^2)\, , \qquad
e^{2\Phi} = \wt g^2 \,k\,  \wh h(\vec {\wh r}, \wh y) \, ,
\ee
\be \label{ex3new}
\wh h(\vec {\wh r}, \wh y) \equiv  {\sqrt k \over 2  \, \wh 
r} \, {\sinh(\wh r/\sqrt k)\over \cosh(\wh r / \sqrt k)
- \cos(\wh y/\sqrt k)}\, , \qquad \wh r \equiv |\vec {\wh r}|\, .
\ee
{}From this we see that for large $k$
the function $\wh h$ is a slowly varying
function of $\vec{\wh r}$ and $\wh y$ near $\wh r=0$, $\wh y=\pi\sqrt k$.
Hence we expect the higher derivative corrections to the solution to
be small in this region and the results given in 
\refb{e6scale} should be reliable.

Given that eq.\refb{e6scale} can be trusted for large $k$, we can now
compare it with \refb{en5scale}. We see that the tension of the
G-type
brane is less than that of the U-type brane, and hence
($\sigma $ invariant) tachyon condensation on the G-type brane
cannot take us to the U-type brane. This is qualitatively
different
from what happens in the absence of NS 5-branes; in that case below a
critical radius the G-type brane becomes heavier than the
U-type brane, and winding tachyon condensation
on
the former takes it to the U-type brane. In fact in
this
case we expect 
the reverse to be true, namely the large radius result that $\sigma$
invariant tachyon condensation on the $y$ wrapped non-BPS 
D(p+1)-brane
takes us to a BPS Dp-brane placed at $y=\pi R$ (plus an anti Dp-brane
with vanishing tension
at $y=0$) should continue to hold all the way to the
small $R$ region. For this we require a $\sigma$
invariant tachyonic mode on the $y$-wrapped D(p+1)-brane. This 
will happen if the total length of the $y$-circle, measured in the string
metric, is large compared to the string length so that the usual
tachyonic kink solution satisfying $T(y)=-T(-y)$ still exists
on this circle. We see from the
metric \refb{ex1} or \refb{ex1new} that this is the case for large 
$k$.\footnote{Naively for any $k$ the length is infinite due to
the infinitely long throat near $y=0$. However we cannot trust
the calculation for finite $k$ due to large $\alpha'$ corrections.}

One can give a physical picture of this situation as follows. For small
$R$ and large $|\vec r|$ where the effect
of the NS 5-branes is small, a Dp-brane 
$\bar {\rm D}$p-brane pair
placed at $y=\pi R$ and $y=0$ respectively has higher tension than
a D(p+1)-brane wrapped along
$y$, and the former can
decay into the latter via tachyon condensation. 
This is the usual situation in the absence of NS 5-branes.
As we approach the
NS 5-branes by reducing
$|\vec r|$,, the tension of the latter remains constant, while the
total tension of the former configuration decreases due to the
decrease in the value of $e^\Phi$. At a certain critical
distance away from the NS 5-branes the former
configuration will have less tension
than the latter configuration, and the 
D(p+1)-brane wrapped along
$y$  becomes unstable against possible decay into
a Dp-brane $\bar {\rm D}$p-brane pair
placed at $y=\pi R$ and $y=0$.

An interesting question is: what happens for low values of $k$?
We shall explore this in \S\ref{s6} and \S\ref{s5}.

\sectiono{Comparison of Different D-branes for Low $k$} 
\label{s6}

So far we have introduced different types of unstable Dp-brane
system in the background of $k$ NS 5-branes of type IIA/IIB
string theory with a transverse circle,
or equivalently in the dual type IIB/IIA string theory on ALF spaces,
and explored how in certain cases tachyon condensation on one type of
brane
can take us to another type of brane.
We now
look for possible relationship between these unstable 
D-branes that does not involve tachyon condensation. 
Of these the relationship between the U and
S type D-branes is easy to comprehend. 
If we begin with a configuration where the NS 5-branes are separated
along the $  y$ direction, then the boundary state of the U-type
D-brane at $\vec {r}=0$, describing
a non-BPS D(p+1)-brane wrapping 
the $ y$ circle, should
coincide with the sum of the boundary states of $k$ different S-type
branes, each connecting a given NS 5-brane to its immediate
neighbour. When the 
NS 5-branes coincide, $(k-1)$ of the S-type D-branes collapse
to zero size in the $  y$ direction whereas the remaining 
D-brane describes a S-type brane stretching all the way around
the $  y$ circle. The collapsed D-branes correspond to tensionless
non-BPS Dp-branes carrying no charge or tension, and
up to these tensionless branes, the U and S type branes can
be identified in this case.
In the dual
ALF geometry this would mean that the boundary state of
a non-BPS D(p+2)-brane
wrapped on a vanishing cycle carrying one unit of gauge field strength
flux through the cycle should coincide with that of a non-BPS
Dp-brane upto addition of boundary states describing 
non-BPS tensionless branes.\footnote{Since a CFT 
whose target space has 
$A_{k-1}$ singularity is singular one may worry about
the meaning of boundary state in this context. We can however make
sense of these statements by beginning at the orbifold point
where half unit of B-flux is switched on through the various
cycles\cite{9507012,9603167} and then 
considering the limit where these fluxes are
turned down to zero.}

Unfortunately, comparison between these and the G-type unstable
Dp-brane is plagued by the lack of
understanding of $\alpha'$ corrections to \refb{e6},
\refb{e6scale}. First of all 
the formul\ae\ \refb{e6} are most likely going to be modified
by $\alpha'$ corrections; there does not seem to be any symmetry
at finite $R$ that protects these results from $\alpha'$ corrections.
This still leaves open the possibility that the zero radius formul\ae\
\refb{e6scale}, which can be argued to be valid for large $k$, may be
exact. In fact this also cannot be strictly true for all $k$;
for $k=1$
the eq.\refb{e6scale} gives a tachyon mass$^2$ less than $-1/2$, 
requiring a negative dimension matter sector
operator for the construction of the corresponding vertex operator.
Since this is not possible in a unitary theory, we expect \refb{e6scale}
to be modified for $k=1$. 
In 
\S\ref{s5} we shall use the description in terms of tachyon effective
potential to suggest a mechanism that could modify the result
for $k=1$ without modifying the results for $k\ge 2$. This will essentially
involve the G-type D-brane for $k=1$
undergoing a second order phase
transition at a finite radius, below which it gets identified with the
U-type D-brane.

If we do assume that eqs.\refb{e6scale}
do not get modified under $\alpha'$ corrections we find that for $k=2$
the tachyon mass$^2$ given in \refb{e6scale} and \refb{en5scale}
are identical\cite{0408073} and furthermore the tensions given in
\refb{e6scale} and \refb{en5scale} also agree.
This is an unexpected result 
since both in the original geometry and
in the dual geometry these D-branes are represented by different
kinds of objects. In the original geometry one is a BPS
Dp-brane placed
at a given point on the circle while the other is a non-BPS
D(p+1)-brane
spread over the circle. In the dual geometry one is a BPS D(p+1) brane
wrapped on the equator of a vanishing 2-cycle, whereas the other can
be represented either as a non-BPS Dp-brane or as a
non-BPS D(p+2)-brane wrapped on the vanishing 2-cycle
with magnetic flux.
However since the string world-sheet theory is strongly coupled in this
case,
neither of these geometric intuitions can be trusted, and
if it turns out that there is an underlying 
non-renormalization theorem for
eq\refb{e6scale}, then it would be a strong indication
that the
G- and the U- type 
branes are identical in this case, 
with the geometric tachyon on one playing the
role of the usual open string tachyon on the 
other.
Since for large radius $R$ of the transverse circle
the G-type D-brane can be considered as the
tachyon kink solution on the U-type brane, the above result
can be interpreted as the merger of the
tachyon kink solution and the trivial
solution describing the original unstable D(p+1)-brane
into a single solution at $R=0$.
 In \S\ref{s5} we shall discuss what
this means for the tachyon effective potential that governs the 
formation of $y$-dependent 
tachyon kink on the U-type brane.

The above result, if correct, would identify the usual non-BPS
Dp-brane placed in an $A_1$ singular geometry to a BPS Dp-brane
placed diametrically across 2 NS 5-branes on a transverse circle
in the dual geometry. This may not sound very exciting since we may
not care about non-BPS Dp-branes in singular spaces; however recall that
the interesting part of dynamics of the open string tachyon 
describing the  tachyon vacuum, the rolling tachyon, or the formation
of the tachyon kink along any of the $p$ spatial directions tangential
to the Dp-brane is universal and 
independent of the background
geometry of the transverse space in which the Dp-brane is 
placed\cite{9911116,0005036,0203211,0203265}. Thus
identifying the open string tachyon with the geometric tachyon in
this system will give a geometric description of most of 
the interesting phenomena involving tachyon condensation on a 
non-BPS Dp-brane in
flat space-time background.

Note that even when the tensions associated with two configurations
become identical, they
could be related by
marginal deformation instead of being identical. 
This is what happens for D-brane systems in the absence of 
NS 5-branes\cite{9808141,9812031}. 
However since
marginal deformation typically changes the spectrum of open strings
on the brane, it would not explain why the tachyon mass$^2$ on the
two types of branes agree. 
In any case existence of a marginal deformation connecting the two
configurations is a highly non-generic situation since it requires
the tachyon potential to develop a flat direction. It is much more
likely for two solutions to merge at a given point in the space of
external parameters, as was demonstrated in a closely related example
in \cite{9906109}.

It is also instructive to compare
the properties of different types of 
branes when the NS 5-branes are
separated from each other. For simplicity we shall consider the case
of $k=2$, with one NS 5-brane at $(\vec{\wt r}=\vec 0, \wt y=c)$ and
the other NS 5-brane at $(\vec{\wt r}=\vec 0, \wt y=
2\pi -c)$.
Since the U-type branes wrap the whole $\wt y$-circle,
their properties are not affected by this
move and  their tensions and tachyon mass$^2$ continue to be
given by \refb{en5scale}.
For the S-type brane, \i.e.\
non-BPS D(p+1)-brane stretched
between the two five branes, we get from \refb{eten1}:
\be \label{eten1xx}
\tau_p''=
{\sqrt 2 }\, \wt g^{-1} \, \TT_p \, \left(1  - {c\over \pi}\right),
\qquad m_T^{\prime\prime 2} = -{1\over 2}\, .
\ee
On the other hand we can calculate the tension and the tachyon
mass$^2$ on a BPS Dp-brane at $\wt y=\pi$ with the help of
eqs.\refb{e3xx}-\refb{e5}, and get, in the scaling limit \refb{elimit}:
\be \label{eten2xx}
\tau_p={\sqrt 2 }\, \wt g^{-1} \, \TT_p \, \cos{c\over 2},
\qquad m_T^{2} = -{1\over 2}(2 -\cos c)\, .
\ee
We
see from \refb{eten2xx} that for $c\ne 0$ we have 
$m_T^2<-{1\over 2}$. Since
such a tachyon will require a negative dimension matter
sector vertex operator this formula certainly needs to be
modified. In \S\ref{s5} we shall suggest a mechanism for this
modification that is similar to the one that modifies the
result for the $k=1$ case.

One might have hoped that study of D-brane boundary states in a closely
related background describing the throat geometry of coincident
5-branes\cite{0406173,0412038}
could shed some light on the possible relationship between the
G and U/S 
type branes described here. However as can be seen from 
\refb{ex1}, \refb{ex3},  the background describing this system for any
given $k$ (say $k=2$) has
no free parameters except the string coupling constant $\wt g$ on which
the world-sheet conformal field theory 
has trivial dependence. Thus we cannot
try to analyze this problem 
with the help of any `nearby' conformal field theory;
we really need to solve the problem exactly in the conformal field
theory of interest.

\sectiono{A Second Order Phase Transition for $k=1$?} \label{s5}

For $k=1$ we have a single NS 5-brane and there is no distinction
between the U and S type branes. Both of them
correspond to an unstable Dp-brane sitting at the center of
Taub-NUT in the dual description. The G-type branes are
however quite mysterious since unlike in the case of multiple
NS 5-branes, in this case the dual geometry does not have a
2-cycle and hence we cannot wrap a BPS D(p+1)-brane on the
equator of the 2-cycle. Thus these must be some new kind
of non-BPS Dp-brane configuration sitting at the center of the Taub-NUT
space.
A natural question is: what happens to this Dp-brane
in the flat space limit?

Since this is an important issue it will be useful to review the
steps leading to this question:
\begin{enumerate}
\item We consider a single NS 5-brane of type IIA/IIB
with a transverse circle, and
place a BPS Dp-brane, with its world-volume parallel to that
of the NS 5-branes, at a diametrically opposite point on this
transverse circle.
All other transverse coordinates of the brane are taken to coincide
with that of the 5-brane. For large radius of the transverse circle
we expect this state to exist since $\alpha'$ corrections
are small. Hence it must also have an appropriate
description in the T-dual Kaluza-Klein monopole background 
in type IIB/IIA theory even
though this background is highly curved in this regime.
\item We now tune the string coupling and the radius of the circle
transverse to the 5-brane to zero keeping their ratio fixed. In the
dual description this corresponds to taking the size of the Kaluza-Klein
monopole to infinity, keeping the string
coupling constant fixed. This
gives rise to flat space-time. The
BPS Dp-brane parallel to the NS 5-brane in the original description
should get mapped to some Dp-brane in type IIB/IIA
in flat space-time in the second
description.  This is the D-brane we want to study.
\end{enumerate}
We can think of several possible scenarios:
\begin{enumerate}
\item It describes a genuinely new unstable Dp-brane in flat space-time
with finite tension and tachyon mass$^2$.
\item $\alpha'$ correction to \refb{e6} drives the 
tension to zero or infinity. 
In this case the `new' Dp-brane does not really exist as an
independent object in the spectrum.
\item In the flat space limit the brane spreads out over the whole
space and does not correspond to a localized Dp-brane with
finite tension.
\item The new non-BPS Dp-brane is in fact identical to the usual non-BPS
Dp-brane, and the formul\ae\ given
in \refb{e6scale} are modified
for $k=1$ to those given in \refb{en5scale}.
This amounts to saying that the G and the U-type
branes become identical in the $R\to 0$ limit.
\end{enumerate}
It is
tempting to speculate that the fourth possibility holds. In that case
we shall have a direct geometric interpretation of the open string
tachyon on a non-BPS D-brane in flat space-time in terms of the
geometric tachyon on a BPS D-brane situated at a point of unstable
equilibrium in a dual geometry. We shall now present some observations
which indicate that this is a likely possibility.
\begin{itemize}
\item Let us express 
the formul\ae\ for the
tension and the tachyon mass$^2$ on this brane given in \refb{e6}
in terms of the variables natural to the dual geometry, 
\i.e.\ $\wt g$ and $\wt R$ 
given in \refb{eg1}. This gives:
\be \label{edd1}
\tau_p = { 2}\,
\wt g^{-1}\, \TT_p \,  (4\wt R^{-2} + 1)^{-1/2}, \qquad
m_T^2 = -{1\over (4\wt R^{-2} + 1)^2}\, .
\ee
Thus at small $\wt R$ where the formula can be trusted, 
$\tau_p$ and $|m_T^2|$ start at small values but begin increasing
as we increase $\wt R$. Thus it is not inconceivable 
that as
we increase $\wt R$ to $\infty$, the values of $\tau_p$ and
$m_T^2$ approach the tension ${ \sqrt 2}\,
\wt g^{-1}\, \TT_p$ and tachyon mass$^2$ $-1/2$ of a usual
non-BPS Dp-brane in flat space-time.

\item Since the geometric tachyon on the G-type brane changes
sign
under the transformation $\sigma$ given in
\refb{ez2}, the corresponding
tachyon
in the T-dual description must also change sign under the image of
$\sigma$ given in \refb{ez2new}. 
The tachyon on the usual
non-BPS Dp-brane in Taub-NUT space
is also also odd under this transformation.
This can be seen either by working in the NS 5-brane background
as discussed in \S\ref{s2}, or directly
in the Taub-NUT geometry due to the
presence of the $(-1)^{F_L}$ factor in \refb{ez2new}. Furthermore,
we have also seen that for large $R$, \i.e.\ small $\wt R$, the
G-type brane can be regarded as a $\sigma$ invariant tachyon
field configuration on the U-type brane. Thus it is natural
to expect that by following this field configuration from small
$\wt R$ to large $\wt R$ 
 the two types of branes can be
related by a $\sigma$ invariant field configuration even for large
$\wt R$.

Now if we take the usual non-BPS Dp-brane, \i.e.\ the
U-type
brane in the $\wt R\to\infty$ limit,
then the requirement of $\sigma$ invariance removes the
tachyonic mode. Thus
all $\sigma$ invariant
field configurations on this brane will have
higher tension than the
tension of the original brane. This would imply that the `new'
Dp-brane must be represented by a classical field configuration
on the usual Dp-brane of positive energy density. Furthermore this
solution must be invariant under the $p+1$ dimensional
Poincare group acting on the Dp-brane world-volume since the
`new' Dp-brane is manifestly invariant under these symmetries.
Such a field configuration would
essentially require a configuration of constant scalar fields. 
While we cannot rule out the existence of such solutions, it will
certainly
be more natural if the `new' Dp-brane 
turned out to be the usual non-BPS Dp-brane.
Note that this argument is special to the $k=1$ case; for $k\ge 2$ the
usual non-BPS Dp-brane is placed in a singular geometry, and there may be
additional $\sigma$ invariant tachyonic mode on this brane which 
could condense and take us to a lower energy density configuration.
Indeed we have argued in \S\ref{s4} by working in the NS 5-brane
description that at least for large $k$ such
$\sigma $ invariant tachyonic modes are present on this brane.
\end{itemize}

We shall now present a concrete analysis using the language of
tachyon effective potential to determine under what condition
we can identify the `new' and the `usual' non-BPS Dp-branes.
Let us begin with a single NS 5-brane with a transverse circle
of radius $R$. At large $R$, 
the G-type D-brane -- a Dp-brane placed
at $y=\pi R$ -- is definitely lighter than
the
U-type D-brane -- a non-BPS D(p+1) brane wrapped along the $y$
direction, and we know that there is a $\sigma$ invariant tachyonic
mode on the latter configuration whose condensation produces
the former configuration. 
In fact since the circle size is large we expect many $\sigma$-invariant
tachyonic modes. We shall assume that the spectrum is discrete.
This may sound unreasonable from the point of view
of the NS 5-brane description since the D(p+1)-brane has an
infinite length along the $y$-direction due to the infinite throat
near the NS 5-brane. However from the point of view of the dual
Taub-NUT geometry we have a Dp-brane sitting at the centre
of a non-singular geometry, and there is no reason why we should
have a continuous spectrum.\footnote{Nevertheless we must admit
that many aspects of this conformal field theory are not
understood and there may be subtle effects which invalidate
our analysis.} As we reduce the value of $R$, the modes of the
tachyon begin acquiring positive contribution to their mass$^2$,
and eventually all $\sigma$-invariant tachyonic modes become
massive in the $R\to 0$ limit since in the dual description we
have a non-BPS Dp-brane sitting at the origin in flat space time
with no $\sigma$-invariant tachyonic mode. We shall begin our 
analysis in a range of values of $R$ where all but one $\sigma$-invariant
tachyonic mode have become massive, and furthermore the
magnitude of the mass$^2$ of this remaining
tachyon is small compared
to the string scale. In this region it should be possible to integrate
out all other modes of the tachyon and define a tachyon effective
potential $V(\phi)$ for this single mode $\phi$. The U-type
brane will correspond to the local maximum of the potential
at $\phi=0$. On the other hand the G-type D-brane 
($\bar{\rm D}$-brane) should be described by some other
local extrema at $\pm\phi_0$ of $V(\phi)$ 
unless they have already merged
with the U-type brane by this time.
If we reduce the value of $R$ further, then below some critical
radius $R_c$ the mode $\phi$ also becomes massive.  
Our goal will be
to explore the fate of the G-type brane during this transition.

$\phi$,
being a mode of the open string tachyon on a non-BPS D-brane,
is odd under $(-1)^{F_L}$. Thus the effective potential $V(\phi)$
must have $\phi\to -\phi$ symmetry.
First let us examine what would happen if $V(\phi)$ were a
quartic potential
of the form
\be \label{ef1}
V(\phi) = {1\over 2} a(R) \phi^2 + {1\over 4} b(R) \phi^4\, ,
\quad b(R)>0\, .
\ee
Now we know that for $R>R_c$, $a(R)$ must be negative
because the field $\phi$ is tachyonic. Hence besides the maximum
at $\phi=0$, \refb{ef1} admits two minimia at $\pm \phi_0$ with
\be \label{eph0}
\phi_0=\sqrt{-a(R)/ b(R)}\, 
\ee
which we can identify as a BPS Dp-brane or $\bar{\rm D}$p-brane
placed as $y=\pi R$.  As we reduce $R$,  $a(R)$ vanishes at the
critical radius $R_c$. 
At this radius 
$\phi_0$ vanishes
and the G and U type branes become identical.\footnote{In
the absence of NS 5-branes the point where $a(R)$ vanishes the whole
potential vanishes and the two types of branes, instead of being
identical, are related by a marginal deformation. However this is a
highly non-generic situation and we are implicitly assuming that
the presence of NS 5-branes turns this into a generic 
situation\cite{9906109}.}
As we decrease $R$ further $a(R)$ becomes positive and
the two solutions
continue to be identical. 
Thus below this 
critical value of $R$ there is no distinction between the G and
U type branes.

Let us now consider the case of a general potential. 
We shall continue to denote by $a(R)$ and $b(R)$ the
coefficients of the quadratic and quartic terms in the potential.
Thus again we have $a(R)<0$ for $R>R_c$ and $a(R)>0$ for
$R<R_c$.
It is easy to see
that a general potential will produce the same results if the following
conditions are satisfied:
\begin{enumerate}
\item For $R>R_c$ the Dp-brane / $\bar{\rm D}$p-brane must
correspond to the minima of the potential closest to the
origin and this feature should continue all the way to the 
critical radius where $a(R)$ vanishes. In other words the potential
should not have any additional extrema corresponding to new
(unstable) D-brane configurations.
\item At the critical radius where $a(R)$ vanishes, $b(R)$ must be
positive. For negative $b(R)$,
 instead of the minima at $\pm \phi_0$ 
merging with
the maxima at $0$, there will be new maxima developing around
$\phi=0$ as $R$ goes below the critical radius.
\end{enumerate}
We see that if either of the above conditions is violated then
there will be new unstable
D-branes in the spectrum of the theory. Thus  the most
economical solution is to have the minima at $\pm\phi_0$
merge with the maximum at $\phi=0$ at the critical point.

If the picture described above is correct then it would seem that
certain
discrete symmetry associated with $\phi\to -\phi$ transformation,
which was broken at the vacua $\phi=\pm\phi_0$, is being restored
below the critical value of $R$. Can we identify this symmetry?
In fact this is just the $(-1)^{F_L}$ symmetry. 
Above the critical value of $R$ the non-BPS D(p+1)-brane
wrapped along $y$, represented by the solution $\phi=0$, is
$(-1)^{F_L}$ invariant, but neither a BPS Dp nor a BPS 
$\bar{\rm D}$p-brane situated at the point $y=\pi R$,
represented by the solutions $\phi=\pm \phi_0$, is
invariant under $(-1)^{F_L}$. If the picture described above is
correct, then it would imply that below the critical
value of $R$  a BPS Dp and a BPS 
$\bar{\rm D}$p-brane situated at the point $y=\pi R$ describe
identical configurations and become $(-1)^{F_L}$ invariant.
This in particular will imply that the Ramond-Ramond part of the
boundary state describing the G-type branes should vanish
below the critical value of $R$.

Note that this analysis also applies to other values of $k$ except that
the potential will have different behaviour in those
cases.\footnote{Since for two or more coincident NS 5-branes the
dual geometry is singular, we cannot apply the previous argument
for the existence of an effective potential. However one can carry
out the analysis by first separating the NS 5-branes in a 
$\sigma$-symmetric
fashion and at the end of the analysis take 
the coincident limit.}
For large $k$ the
analysis of \S\ref{s4} shows that $a(R)$ remains negative as
$R\to 0$ and hence the solutions describing the two types of
Dp-branes remain distinct. If we believe that eqs.\refb{e6scale}
are not renormalized then this should continue to hold till $k=3$,
so that the G-type brane always remains lighter than the
U-type brane. On the other hand for $k=2$, $a(R)$ should vanish
precisely at $R=0$ so that the two types of branes become
identical at that point.

This analysis also suggests a mechanism that would modify the
results \refb{e6scale} for $k=1$ without modifying them for $k\ge 2$.
This happens essentially because for $k=1$ the G-type D-brane
undergoes a second order phase transition at the critical radius $R_c$.
For $R<R_c$ these D-branes merge with the U-type D-branes
and hence the relevant formul\ae\ to use are those in \refb{en5scale}.
A naive analytic continuation of the results from the
$R>R_c$ region will give us the corresponding quantities for
the unphysical (complex) solutions. The same reasoning should
apply for the unphysical answers gotten in \refb{eten2xx} for the
G-type D-brane in the presence of a pair of separated NS 5-branes.
Since at zero separation $c$ the branches describing the G and
U/S type D-branes meet exactly at $R=0$, it is quite
likely that for non-zero $c$ the branches will meet at non-zero $R$
and the G-type D-brane will encounter a second order phase
transition where it merges with the S-type D-brane.
Below this critical radius the relevant formul\ae\ are those given
in \refb{eten1xx}. 
Thus in the $(R,c)$ plane there will be a line of second order fixed
points passing through the $(R=0, c=0)$ point. 
In fact this phase transition 
is not disconnected from the
one we encountered in the $k=1$ case. For $c=\pi/2$ the configuration
of the pair of NS 5-branes is just the double cover of the $k=1$
configuration, and the G and S type D-branes sees background
identical to the G and U/S type D-branes in the $k=1$
case.  
Thus at $c=\pi/2$ the line of critical point in the $(R,c)$
plane should reach the $k=1$
critical point $R=2 R_c$, the factor of 2 accounting for the fact that
the configuration for $k=2$ is the double cover of the configuration
for $k=1$. The same reasoning can also be applied to the case of
$k$ NS 5-branes. In this case we should have a codimension one
critical surface in the full moduli space labelled by the radius $R$
and the possible separation of the NS 5-branes maintaining the $\sigma$
symmetry. This critical curve does not pass through any physical value
of $R$ when all the branes are coincident, but passes through the point
$R=k R_c$ when the $k$ NS 5-branes are situated as equal intervals
$2\pi R/k$ along the transverse circle. 

If the non-renormalization theorems are violated, then the
precise details of the critical surface will be different from
the one given above. However we would expect the general features
to remain, assuming that the quartic term in the tachyon effective
potential for $k=1$ has the correct sign. In this case we can start
at the critical point corresponding to $R=k R_c$ and equally separated
NS 5-branes, and follow its fate in the moduli space of NS 5-brane
configurations.

If it turns out that the non-renormalization theorems do hold, 
and the identification of the non-BPS
Dp-branes with geometrically unstable Dp-branes in the T-dual theory
hold both for two coincident NS 5-branes as well as a single
NS 5-brane at $R=0$, then we have two possible
description of the open string tachyon on a non-BPS D-brane as a
geometric tachyon. Which one is more useful?
Our first thought may be that the $k=1$ case is more useful because
it lands us directly on non-BPS Dp-branes in flat space-time.
The reverse however is true. For the $k=1$ case we have seen that
even if our picture of branch merger holds, there is a switch of branch
for the geometrically
unstable D-branes at the critical radius. In particular if we
had made a naive analytic continuation of various physical quantities
of this system
to values of $R$ below the critical value, we would get the
wrong answer because it will land us into the wrong branch of 
(possibly complex) solutions which do not correspond to any physical
D-brane. On the other hand if for $k=2$ the two branches precisely
meet at $R=0$, then all the universal properties of the 
tachyon on the non-BPS 
Dp-brane
can be derived by analytic continuation of the results for the
geometrically unstable Dp-brane to $R=0$.

One aspect of this correspondence may seem puzzling.
One might  wonder how the boundary state of a Dp-brane that is
localized at $y=\pi R$ could coincide with that of a D(p+1)-brane
that spreads out along the $y$ circle. This however is not a
serious problem since the tidal forces on the Dp-brane
due to the $y$ dependent
dilaton will tend to spread out the boundary state away from
$y=0$. Indeed this has been observed even in simpler cases
of hairpin brane boundary state\cite{0310024,0312168} 
where we have a linear
dilaton background.

\sectiono{Discussion} \label{s7}

In this paper we have argued that under certain conditions a
BPS D-brane with geometric instability   due to being placed
at a saddle point of the potential may be
identified with a non-BPS D-brane with the usual tachyonic
instability. This would give a geometric interpretation of the
usual open string tachyons. It will be interesting to see if
this geometric picture can provide some insight into the
analytic solutions of string field equations describing the tachyon
vacuum and various solitons, -- a problem whose bosonic
counterpart has only been solved
recently\cite{0511286,0603159,0603195,0605254,0606131,0606142,
0611110,0611200,0612050}.
Another aspect of tachyon condensation where the present analysis
may throw some light is vacuum string field 
theory\cite{0012251}, -- string field theory around the tachyon vacuum. 
For a
geometrically unstable D-brane the tachyon vacuum represents a
D-brane sitting at the core of an NS 5-brane. This may provide some
insight into the nature of the open string tachyon vacuum.

\bigskip

{\bf Acknowledgement:}
I would like to thank David Kutasov for useful communications
and Barton Zwiebach for  useful discussions and
comments on an earlier version
of the manuscript.
This work was supported  generously
by the people of India, J.C. Bose fellowship of the
Department of Science and Technology of Govt. 
of India and Moore distinguished scholarship at
California Institute of Technology.

\end{document}